# Modeling Long Memory in REITs

John Cotter, University College Dublin[*]

Centre for Financial Markets, School of Business, University College Dublin, Blackrock, County Dublin, Republic of Ireland. E-Mail: john.cotter@ucd.ie

and

Simon Stevenson, Cass Business School, City University

Faculty of Finance, Cass Business School, City University, 106 Bunhill Row, London, EC1Y 8TZ, UK. Tel: +44-20-7040-5215, Fax: +44-20-7040-8881, E-Mail: s.stevenson-2@city.ac.uk

Current Draft: November 2006

---

[*] Author to whom correspondence should be addressed.

# Modeling Long Memory in REITs

## Abstract

One stylized feature of financial volatility impacting the modeling process is long memory. This paper examines long memory for alternative risk measures, observed absolute and squared returns for Daily REITs and compares the findings for a non-REIT equity index. The paper utilizes a variety of tests for long memory finding evidence that REIT volatility does display persistence, in contrast to the actual return series. Trading volume is found to be strongly associated with long memory. The results do however suggest differences in the findings with regard to REITs in comparison to the broader equity sector which may be due to relatively thin trading during the sample period.

# Modeling Long Memory in REITs

## 1. Introduction

The continued development and increase investor awareness of the Real Estate Investment Trusts (REIT) sector has led in recent years to a dramatic increase in daily trading in the sector. SNL Financial estimate that average daily volume has increase from just over 7m in 1996 to over 40m in 2005. In addition, as the sector continues to mature and develop there will be increased interest in derivative products based on the sector. At present a number of OTC (over-the-counter) products are available, while the Chicago Board Options Exchange (CBOE) provides traded options on the Dow Jones Equity REIT Index. The growth in both traded and OTC derivative products based on REITs furthers the interest in the dynamics of the sector at higher frequencies such as daily intervals. Furthermore, an examination of the volatility of the sector becomes more important as not only daily trading increases but also due to the role of volatility in derivative pricing.

In contrast to the large literature that has examined the return behavior of REITs; very few have examined volatility in the sector, with even an even smaller number to have done so using high frequency data. Two early papers on REIT volatility (Devaney, 2001 and Stevenson, 2002b) both analyzed monthly data. The analysis conducted by Devaney (2001) was primarily concerned with the sensitivity of REIT returns and volatility to interest rates and was undertaken in a GARCH-M framework. The Stevenson (2002b) paper was, in contrast, concerned with the volatility spillovers present in monthly data within both different REIT sectors and between REITs and the equity and fixed-income markets. Four recent papers have examined various aspects of daily REIT volatility. Winniford (2003) concentrates on seasonality in REIT volatility. The author finds strong evidence that volatility in Equity REITs varies on a seasonal basis, with observed increased volatility in April, June, September, October and November. Cotter & Stevenson (2006) utilize a multivariate GARCH model to analyze dynamics in REIT volatility. Using a relatively short and quite distinct period of study (1999-2003) they find an increasing relationship between Equity REITs and mainstream equities in terms of both return and volatility. Bredin et al. (2005), as with the Devaney (2001) paper, concentrate on the specific issue of interest rate sensitivity, examining the impact of unanticipated changes in the Fed Funds Rate on REIT volatility. The final paper to have examined daily REIT volatility is the one most similar to the current paper. Najand & Lin (2004) utilize both GARCH and GARCH-M models in their analysis, reporting that volatility shocks are persistent.

This persistence in volatility is a common empirical finding in financial economics and was studied extensively in Taylor (1986). Whereas asset returns have largely been found to contain very little autocorrelation, it has been noted in a large number of papers across different asset classes that autocorrelation in various measures of volatility does exist at significant levels and remains over a large number of lags[1]. This effect, referred to as long memory has been documented across a large sphere of the finance literature from macroeconomic series such as GNP (Diebold & Rudebusch, 1989) to exchange rate series (Dacorogna et al., 1993; Baillie et al., 1996; Andersen & Bollerslev, 1997a, 1997b) at low and relatively high frequencies. Moreover, it is documented for equity index series at daily intervals (Ding et al., 1993, Ding & Granger, 1996).

This paper examines the long memory properties of alternative risk measures, observed absolute and squared returns for REITs and compares these to the non-REIT S&P500 composite index. Analysis of long memory has been overlooked and we benchmark our REIT findings against the broader equity market. Specifically, the long memory property and its characteristics are explored. The long memory property occurs where volatility persistence remains at large lags and the series are fractionally integrated. Much focus has been on the absolute returns series, $|R_t|^k$, or a squared returns series, $[R_t]^k$, for different power transformations, $k>0$. This property adds to the general clustering condition usually referred to in the context of squared returns persistence originally modeled in Engle's (1982) ARCH paper. Daily seasonality of the dependence structure shows a slow decay of the autocorrelation structure involving a *u*-shaped cyclical pattern (Andersen & Bollerslev, 1997a, 1997b). In addition, Ding et al. (1993) indicate that this non-linear dependence is strongest for absolute volatility with a power transformation of $k=1$ suggesting that modeling should focus at the level of returns rather than squared returns in building parametric econometric volatility specifications.

This paper begins by examining the autocorrelation structure of the REIT returns and volatility series. It then formally tests for the long memory property and measures the magnitude of the fractional integration parameter. In terms of model building, there are several approaches from linear and non-linear perspectives that could be applied. This paper fits two long memory volatility models, Fractionally Integrated GARCH (FIGARCH) and (FIEGARCH) that allow for asymmetry. Baillie et al. (1996) find that these models have considerable success in modeling daily equity returns and we will investigate whether these GARCH models can capture the long memory properties of daily REITS. The paper proceeds as follows. In Section 2, long memory is discussed. The section is completed by a presentation and discussion of our GARCH models that are fitted to the daily series'. Details of the series and data capture follow in Section 3. Section 4 presents the empirical findings. It begins by briefly describing the indicative statistics of the volatility series', followed by a thorough analysis of their long memory characteristics. In addition, the ability of the GARCH processes to model volatility persistence is presented. Finally, a summary of the paper and some conclusions are given in section 5.

## 2. Long Memory

Baillie (1996) shows that long memory processes have the attribute of having very strong autocorrelation persistence before differencing, and thereby being non-stationary, whereas the first differenced series does not demonstrate persistence and is stationary. However, the long memory property of these price series is not evident from just first differencing alone, but has resulted from analysis of risk measures. In fact financial returns themselves have only been found to exhibit short memory, with significant first order dependence that dissipates rapidly over subsequent lags. Thus the finance literature has concentrated its analysis of long memory on the volatility series and we follow this convention.

Long memory properties may be investigated by focusing on the absolute returns series, $|R_t|^k$, or the squared returns series denoted squared volatility, $[R_t^2]^k$, and on their power transformations, where k>0[2]. The former are examined as Davidian & Carroll (1987) find that absolute realizations are more robust to the presence of fat-tailed observations found in financial series than their squared counterparts whereas the latter are utilized given their popularity in underpinning the commonly used risk measures such as standard deviation and variance. Moreover, empirical analysis of financial time series suggests that the long memory feature dominate for absolute over squared realizations (see Ding & Granger, 1996).

Models with a long memory property have dependency between observations of a variable for a large number of lags so that $Cov[R_{t+h}, R_{t-j}, j≥0]$ does not tend to zero as $h$ gets large[3]. In particular, long memory in financial time series has concentrated on volatility realizations where unexpected shocks affect the series for a large time frame. Thus confirmation of long memory properties for REITS would have major implications for the associated investments strategies that need to take account of the persistence and characteristics of the dependence structure in REIT volatility. However, if the dependency between observations of a variable disappears for a small number of lags, $h$, such as for a stationary ARMA process, then the data is described as having a short memory property and $Cov[R_{t+h}, R_{t-j}, j≥0] \to 0$. Formally, long memory is defined for a weakly stationary process if its autocorrelation function $\rho(\cdot)$ has a hyperbolic decay structure:

$$\rho(j) \sim C_j^{2d-1} \text{ as } j \to \infty, C \neq 0, 0 < d < \frac{1}{2} \tag{1}$$

In contrast, short memory, or anti-persistence is evident if $-1/2 < d < 0$.

The corresponding shape of the autocorrelation function for a long memory process is hyperbolic if there is a relatively high degree of persistence in the first lags that declines rapidly initially and that is followed by a slower decline over subsequent lags but the length of decay remains strong for a very large number of time periods. Previous analysis of equity returns suggest that the long memory parameter or degree of fractional integration, $d$, is generally found to be between 0.3 and 0.4 (e.g. Andersen and Bollerslev, 1997a; and Taylor, 2000).

The explanations for long memory are varied. One economic rationale results from the aggregation of a cross-section of time series with different persistence levels (Andersen & Bollerslev, 1997a; Lobato & Savin, 1998). Alternatively, regime switching may induce long memory into the autocorrelation function through the impact of different news arrivals (Breidt et al., 1998). The corresponding shape of the autocorrelation function is hyperbolic, beginning with a high degree of persistence that reduces rapidly over a few lags, but that slows down considerably for subsequent lags to such an extent that the length of decay remains strong for a large number of time periods. Also, with a slight variation, it may follow a slowly declining shape incorporating cycles that correspond to, for example, daily seasonality (Darcorogna, et al., 1993).

We test for the existence of long memory in REITs by using an informal analysis of autocorrelation dependence of our returns and volatility series augmented by two formal tests for the existence of the property. We are interested in two issues: whether REITs exhibit long memory properties and how the characteristics of the dependence structure of REITs compares to non-REIT equities. The first test statistic is the parametric Modified Rescale Range (*R/S*) statistic developed by Lo (1991):

$$Q_T = \frac{1}{\hat{\sigma}_T(q)} \left[ \max_{1 \leq k \leq T} \sum_{j=1}^{k} (z_j - \bar{z}) - \min_{1 \leq k \leq T} \sum_{j=1}^{k} (z_j - \bar{z}) \right] \quad (2)$$

Where $\hat{\sigma}_T$ is the estimate of the long run variance and for any series z, we compare the realized value $z_j$ to its mean, $\bar{z}$ and examine the range of the variation. The Modified *R/S* allows for short memory in the time series but can distinguish if long memory exists separately, whereas in contrast, the original *R/S* statistic (Hurst, 1951) is not able to distinguish between long and short memory. Given, that microstructure issues such as bid-ask bounce induces first order correlation and short memory in returns series (Andersen et al., 2001) we may have both long and short memory characteristics in the series analyzed[4]. As a by product, we can also obtain an estimate of the degree of fractional integration, *d*, from applying this test denoted *R/S d*. This describes the degree of fractional integration and allows us compare to benchmarks, for example whether it is in the domain $0 < d < \frac{1}{2}$, and its magnitude across different REIT and non-REIT series. We ensure robustness of the approach by ensuring an adequate number of lags are analyzed by applying the least absolute deviation in fitting the test statistic.

In addition, long memory is investigated by using the semi-nonparametric Geweke & Porter-Hudak (1983) log-periodogram regression approach (*GPH*) updated for non-Gaussian volatility estimates by Deo & Hurvich (2000). This adjustment is required given the fat-tailed and skewed behavior of financial time series. We also obtain semi-non parametric estimates of the long memory parameter denoted *GPHd*. Assuming, $I(\omega_j)$ stands for the sample periodogram at the $j^{th}$ fourier frequency, $\omega_j = 2\pi j/T$, $j=1, 2, \ldots, [T/2]$, the log-periodogram estimator of $d_{GPH}$ is based on regressing the logarithm of the periodogram estimate of the spectral density against the logarithm of $\omega$ over a range of frequencies $\omega$:

$$\log[I(\omega_j)] = \beta_0 + \beta_1 \log(\varpi_j) + U_j \tag{3}$$

where $j=1, 2,\ldots, m$, and $d=-1/2\beta_1$. This approach allows us to determine if the long memory property is evident in the series analyzed and gives estimates of the long memory parameter. Again like the *R/S* approach, estimates of *d* are dependent on the choice of *m*. We estimate the test statistic by using $m=T^{4/5}$ as suggested by Andersen et al. (2001). This implies that for our sample size, a sample of 788 periodogram estimates is employed in our analysis.

Given, that long memory is not evident in financial returns series, but is strongly found in their volatility counterparts we need to examine volatility models and their suitability in describing the persistence patterns of the REIT and non-REIT series. Whilst second order dependence is a characteristic of financial returns, usually modeled by a stationary GARCH process, these specifications have been questioned as to their ability to model the long memory property adequately in contrast to their Fractionally Integrated GARCH counterparts (Baillie, 1996). For instance, while stationary GARCH models show the long memory property of financial returns volatility series occurs by having $[R_t^2]$ and $|R_t|$ with strong persistence, they assume that the autocorrelation function follows an exponential pattern not corresponding to a long memory process. In particular, the correlation between $[R_t^2]$ and $|R_t|$ from stationary GARCH models and their power transformations remain strong for a large number of lags, with the rate of decline following a constant pattern (Ding et al., 1993), or an exponential shape (Ding & Granger, 1996). In contrast, a number of returns series, both $[R_t^2]$ and $|R_t|$, in fact have been found to decay in a hyperbolic manner, namely, they decline rapidly initially, and this is followed by a very slow decline (Ding & Granger, 1996)[5].

Turning to the set of conditional volatility models applied in this study, we first use the Fractionally Integrated GARCH (FIGARCH) model introduced by Baillie et al. (1996). These incorporate the standard time-varying volatility models and estimate the short run dynamics of a GARCH process. More importantly, they also measure the long memory characteristic of the data by estimating the degree of fractional integration *d*. First, taking a GARCH (*p*,*q*) process.

$$\sigma_t^2 = a_0 + \sum_{i=1}^{p} a_i \varepsilon_{t-i}^2 + \sum_{i=1}^{p} b_j \sigma_{t-j}^2 \qquad (4)$$

It can be written as an ARMA (m, q) process

$$\phi(L)\varepsilon_t^2 = a + b(L) \qquad (5)$$

For

$$\mu_t = \varepsilon_t^2 - \sigma_t^2 \qquad (6)$$

$$\phi(L)1 - \phi_1 L - \phi_2 L^2 - ..... - \phi_m L^m \qquad (7)$$

$$b(L)1 - b_1 L - b_2 L^2 - ..... - b_q L^q \qquad (8)$$

Converting it back into a GARCH type process gives the FIGARCH (*m*,*d*,*q*) model:

$$b(L)\sigma_t^2 = a + [b(L) - \phi(L)(1-L)^d]\varepsilon_t^2 \qquad (9)$$

This model can be expanded to deal with further stylized features of financial data. For instance, leverage effects have been found (Black, 1976). Thus volatility is affected asymmetrically with bad news having a greater impact than positive news leading to the introduction of the (Exponential) EGARCH process. If the effects of news are long lasting as suggested by the fractionally integrated process we should also determine if the long memory exhibits asymmetric effects. In order to models this we also apply the Exponential version of the FIGARCH, the FIEGARCH developed by Bollerslev & Mikkelsson (1996):

$$\phi(L)(1-L)^d \ln \sigma_t^2 = a + \sum_{j=1}^{q} b_j |x_{t-j}| + \gamma_j x \qquad (10)$$

The residuals from both processes, $\varepsilon_t$, were initially assumed to be from a conditionally fat-tailed process in line with the commonly found characteristics in financial returns. We assumed that the underlying data conditionally followed a student-*t* distribution as in Baillie & DeGennaro (1990).

**3. Data**

The data used in this paper consists of daily logarithmic returns for the period January 1 1990 through December 30 2005 totaling 4175 observations. During this time the popularity of REITS has expanded dramatically with massive growth in investor awareness and interest focusing in on the return and volatility characteristics of the sector. As we are interested in the long memory of the REIT sector we compare the findings to the general equity markets, as represented by the S&P 500 Composite.

Some descriptive statistics of the respective series are outlined in Table 1 detailing the first four moments of each series and a test for normality. Separate analysis is completed for the returns series and the two proxies of volatility, absolute and squared volatility. Starting with returns we find that the average daily returns of both series are near zero but positive for the time frame analyzed suggesting that for the mainstream equity market the 1990s boom has slightly outweighed the downturn at the start of this decade. Accordingly, the reverse is true for the REIT sector, with their strong recent performance outweighing the underperformance of the sector observed during the late nineties. The average risk of REITs approaches 1% and is almost identical to the overall market as proxied by the S&P.

The relative tranquility of the REIT series can easily be seen in the time series plots given in Figure 1. Here we can see the increase in volatility at the turn of the decade associated with amongst other events, the fall out of the Asian crises and September 11, and the technology bubble where equity markets in general exhibited greater turbulence and very poor return performance. In the last couple of years the markets have settled down to some degree. Evidence on higher moments suggests negative skewness recorded by all series suggesting that the weights of the large negative returns are dominating their positive counterparts. Consistent with the literature, we also find excess kurtosis suggesting that the series exhibit a fat-tailed property. Combining these findings for skewness and kurtosis, we find that all series are non-normal using the Jarque-Bera test statistic and therefore need to incorporate this property later in our modeling approach.

Turning to the proxies of the volatility series, we first reiterate the findings for the returns series, namely, that the REIT index behaves very similarly to the S&P in this context. But by looking at the time series properties of these volatility measures directly we also get some further insights. Looking at the plots in Figure 1 we see the relative magnitude of the volatility. Clearly REITs are less volatile than the general market and as as a time-series they exhibits smaller deviations, especially if you look at the squared realizations. We also see the volatility clustering property where periods of high volatility or low volatility can remain persistent for some time before switching. This property suggests that volatility on any day is dependent on previous day's values and we will model this phenomena using a GARCH process that specifically incorporates long memory. The lack of independence of either absolute or squared volatility is clearly seen by the lack of normality and excess kurtosis reported in Table 1 for both series. As volatility is positive, we get strong positive skewness for all series that is reasonably similar across the series. Comparing the two measures of volatility, we see that the magnitude of the squared realizations dominate their absolute counterparts but that the squared values are more prone to extreme outliers regardless of which series you examine.

## 4. Empirical Analysis

Our main focus in this paper is to examine the long memory properties of REITs and it is to this issue that we now turn. We begin by discussing the autocorrelation plots; followed by formal testing for long memory and determining the magnitude of a long memory parameter, and finally we outline our findings from applying two time-varying long memory volatility models. First, looking at dependence using the autocorrelation function (ACF), we provide plots over 100 lags for the returns series given in Figure 2 (as we will see our findings are reiterated from the more formal testing that we will discuss later). We repeat this analysis for our volatility series and these are given in Figure 3 for absolute volatility and Figure 4 for squared volatility. First, looking at returns, we see that clearly that there is a lack of dependence across nearly all lags for the REIT and non REIT series. Thus returns do not have a long memory property. This is in line with previous findings for financial returns where series are considered to be near white noise or independently distributed. There are some exceptions and one of note is the strong positive autocorrelation exhibited for the first lag of the REIT index indicating that, on average, positive (negative) returns tend to be followed over the next day by positive (negative) returns. In contrast, there is a lack of supporting evidence for the magnitude and pattern of subsequent lags. Given that the results are based across the entire sample period these findings for REITs may be influenced by earlier observations at a point when substantially less daily trading occurred in the sector. Thin trading is a traditional explanation for first order autocorrelation (Scholes and Williams, 1977).

We now turn to our volatility series. As Ding et al. (1993) suggest that, as volatility is unobservable, the long memory in equity data should be examined for different power transformations of the volatility proxy series. We follow this suggestion by examining the volatility series for 5 different power transformations [$k$=0.25, 0.5, 1, 1.5, 2]. This supports the analysis of Beran (1994) in his seminal work in the area. In its strictest sense, the ACF plots in figures 3 and 4 do not offer conclusive evidence that REITs exhibit long memory in volatility but are much more striking in their support for the property in the non REIT indexes. Moreover there is strong variation in the strength of the long memory feature for the different power transformations and it tends to be stronger for lower *k*. These findings are consistent for squared and absolute volatility. It is noticeable that REITs appear to display less persistence in volatility than the general market. The ACF plots for the S&P indexes report enhanced long memory. It can be seen that in general the first lag for the REIT volatility ACF's tends to be of a greater magnitude but that the persistence reduces at a faster rate than for mainstream equities. This may again be due to the relative degrees of daily trading in the sector, with perhaps thin and non-synchronous trading leading to an enhanced short-term impact.

Table 2 reports details of the initial tests for long memory using the approaches described in Section 2. There is no evidence of long memory in relation to the return series analyzed. However, in contrast, there is extensive evidence of long memory in both the absolute and squared volatility series'. This is consistent across all of the different power transformations, although the effect is generally enhanced as *k* reduces, particularly in the case of REITs. Furthermore, generally the magnitude of the test statistics is lower for the REIT sector than for the S&P. As with the ACF plots this may be the result of non-synchronous or thin trading in the REIT sector, particularly earlier in the sample period. The results for the GARCH based tests are contained in Table 3. The results generally show that both the FIGARCH and FIEGARCH models provide good fits for the data, furthermore the results are broadly in line with expectations and the previously reported findings. The degree of fractional integration, as measured by the d-values, is in the range of 0.3-0.4 for the FIGARCH model for both series and is consistent with the previous empirical evidence. In relation to the FIEGARCH model the significant leverage coefficient also implies asymmetry in the long memory process with the impact of negative shocks having a proportionally larger impact not only on immediate volatility but also the persistence present.

In the literature volume is seen as an important explanatory variable for time varying volatility[6]. To investigate whether trading volume is important for the long memory inherent in the volatility series we analyse the role of trading volume. In Figure 5 we see that large increase in trading activity in equities and this is particularly pronounced for REITs that had very low volume at the start of the sample. In Figure 6 we see that the change in trading volume shows similar patterns to that of the price series, namely, there is clustering of inactive (active) trading periods followed by active (inactive) trading periods.

Taking the volume data we fit a FIEGARCH model and results are reported in Table 4 with the associated time series plots given in Figure 7.[7] We are trying to determine whether volatility is an important mixing variable for long memory in volatility. Trading volume is clearly an important explanatory variable for our conditional volatility with a strong statistical significance. Also, economically a 1% change in REIT volume causes a 0.01% change in its volatility and this effect is approximately doubled for the S&P series. Interestingly, by including the change in volume variable we see a major revision in the volatility specification with GARCH and ARCH coefficients being considerably amended in comparison to the FIEGARCH model results excluding volume. The main coefficients of the GARCH process, whilst remaining significant, provide support for the hypothesis that volume and volatility are strongly related. The impact of volume, however, is even more pronounced on long memory with the long memory parameter, d, increasing to approximately 0.8 for both REITs and S&P series. Thus the long memory characteristic is no longer present in the volatility series if we include trading volume as an explanatory variable. Overall, changes (increases) in volume are strongly associated with the long memory in property found in REIT (and non-REIT) data.

## 5. Conclusion

This paper has examined the long memory properties in the volatility of the REIT sector at daily frequencies. As the sector develops and daily trading volume increases not only will interest in the daily dynamics in REITs increase but it will also in all likelihood increase interest in derivative instruments based on the sector. The paper illustrates that as with the general equity market volatility persistence occurs in contrast to a lack of persistence in the return series. However, there is evidence that long memory in REIT volatility is not of the same magnitude as that observed in the S&P 500 index. Moreover, changes in volume is an important explanatory variable in modeling long memory of REIT volatility.

Tables & Figures

**Figure 1: Time Series Plots of Daily Series**

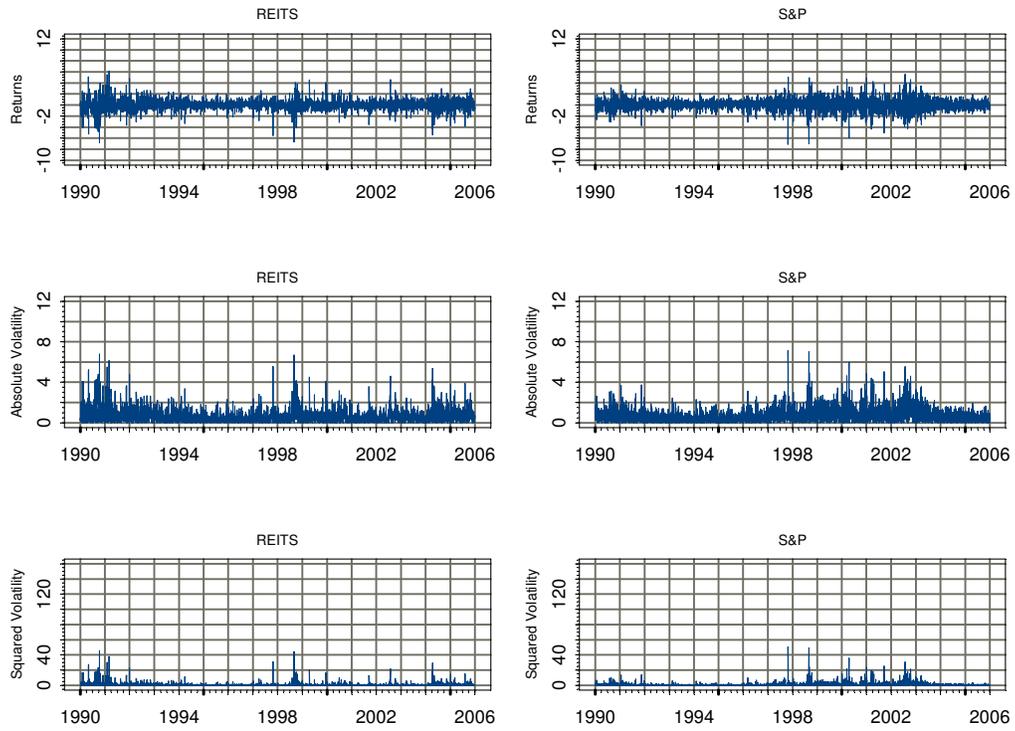

**Table 1: Summary Statistics for Daily Series**

|  | REITs | S&P 500 |
|---|---|---|
| **Panel A: Returns** | | |
| Mean | 0.02875 | 0.03023 |
| Std Dev | 0.9444 | 0.9965 |
| Skewness | -0.2562 | -0.1004 |
| Kurtosis | 8.297 | 7.011 |
| Normality | 4925.84 | 2805.12 |
| **Panel B: Absolute Volatility** | | |
| Mean | 0.6591 | 0.701 |
| Std Dev | 0.677 | 0.7088 |
| Skewness | 2.617 | 2.244 |
| Kurtosis | 14.55 | 11.66 |
| Normality | 27963 | 16541.8 |
| **Panel C: Squared Volatility** | | |
| Mean | 0.8926 | 0.9936 |
| Std Dev | 2.405 | 2.432 |
| Skewness | 8.535 | 8.396 |
| Kurtosis | 109 | 119 |
| Normality | 2003006 | 2387327 |

Notes: Mean and standard deviations are expressed in percentage form. Normality is tested for by the Jarque-Bera test statistic.

**Figure 2: Plots of Autocorrelation Values for Daily Returns Series.**

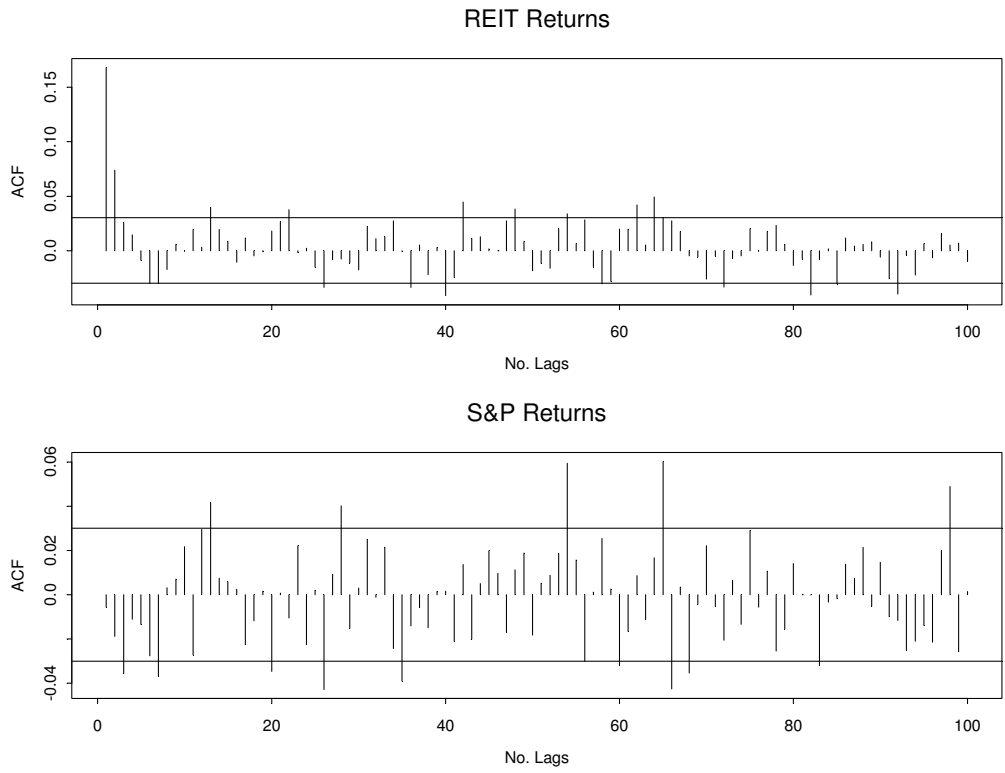

Notes: Confidence intervals are imposed on each plot using the critical values for each series.

**Figure 3a: Plots of Autocorrelation Values for REIT Daily Absolute Volatility**

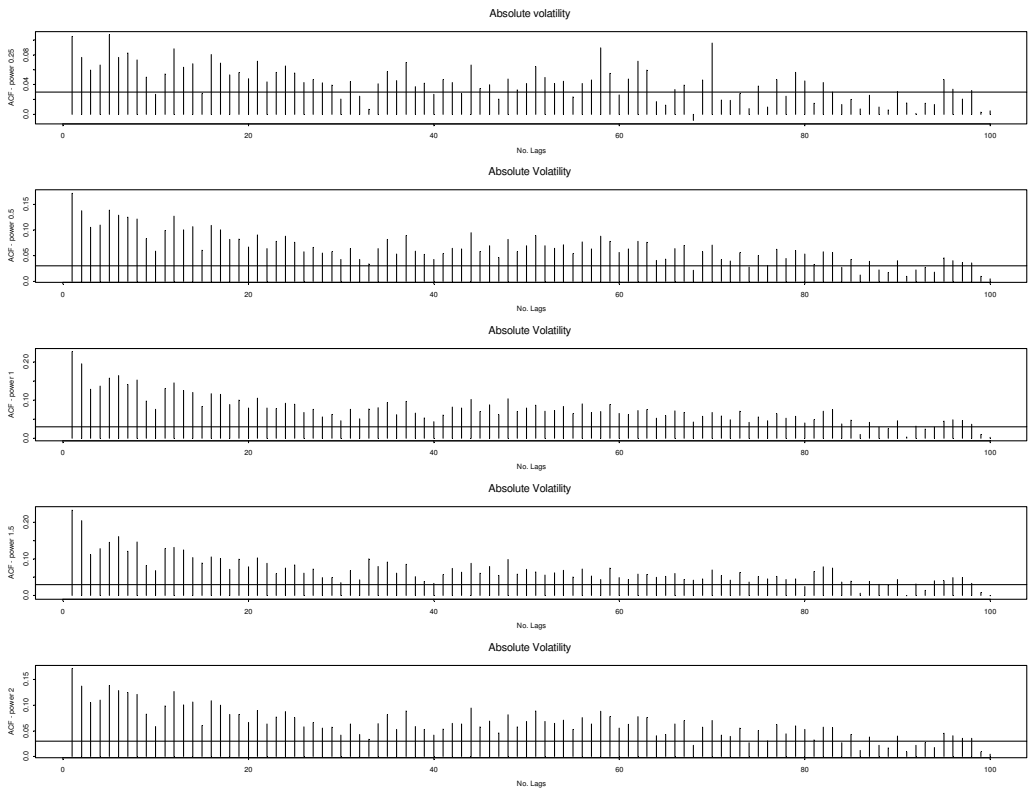

Notes: Confidence intervals are imposed on each plot using the critical values for each series.

**Figure 3b: Plots of Autocorrelation Values for S&P Daily Absolute Volatility**

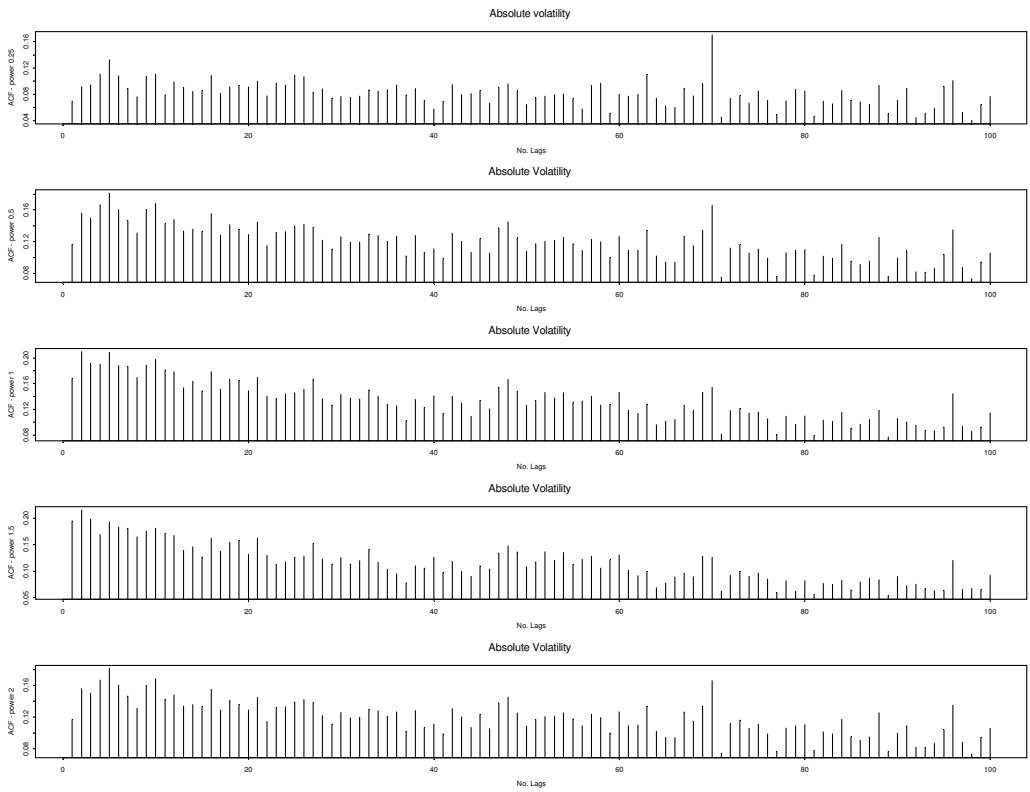

Notes: Confidence intervals are imposed on each plot using the critical values for each series.

**Figure 4a: Plots of Autocorrelation Values for REIT Daily Squared Volatility**

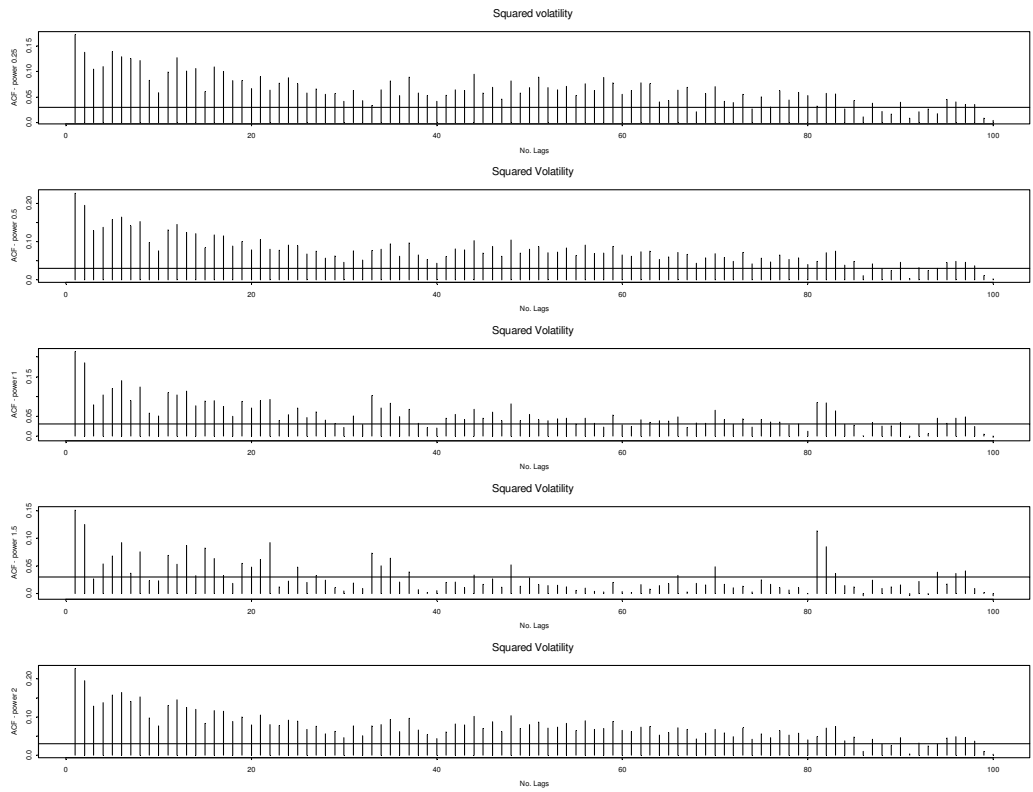

Notes: Confidence intervals are imposed on each plot using the critical values for each series.

**Figure 4b: Plots of Autocorrelation Values for S&P Daily Squared Volatility**

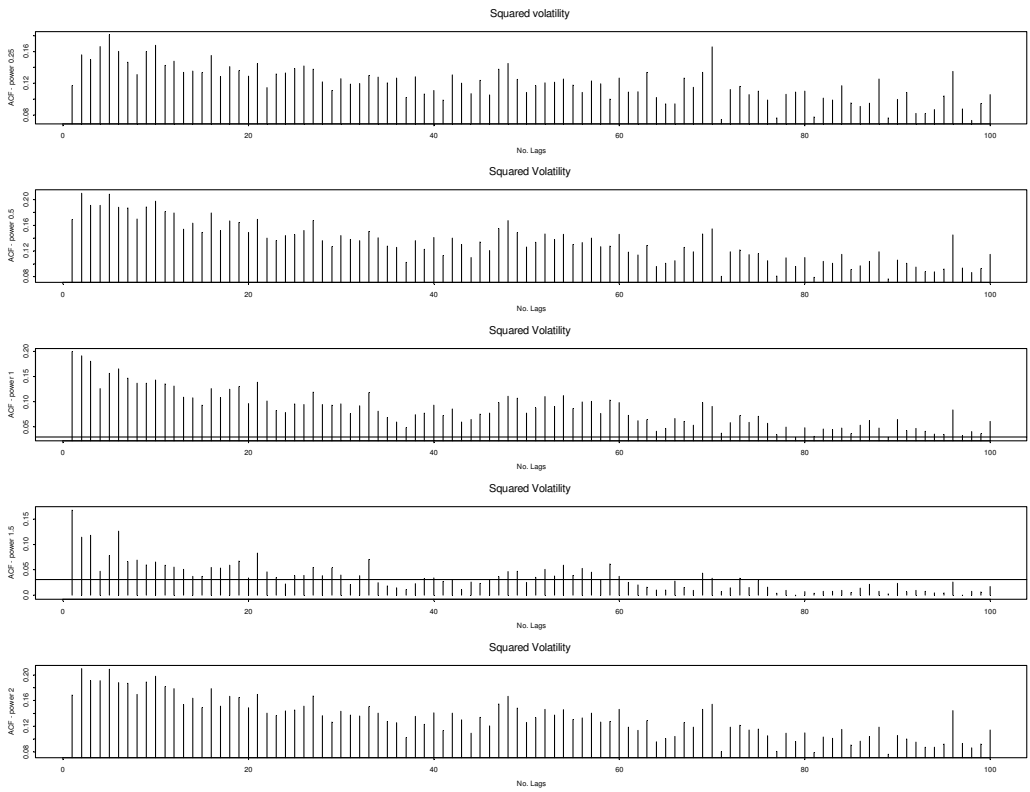

Notes: Confidence intervals are imposed on each plot using the critical values for each series.

**Table 2: Long Memory Diagnostics for Daily Series**

|  |  | R/S | GPH | R/S d | Periodogram d |
|---|---|---|---|---|---|
| **Panel A: Returns** | | | | | |
| REIT | | 1.577 | -0.0715 | 0.141555 | -0.0155121 |
| S&P | | 1.5257 | 0.5438 | 0.087804 | -0.0225327 |
| **Panel B: Absolute Volatility** | | | | | |
| REIT | k = 0.25 | 4.1083** | 3.703** | 0.139796 | 0.2791813 |
|  | 0.5 | 4.2062** | 4.0009** | 0.162991 | 0.3362795 |
|  | 1 | 3.8566** | 4.5319** | 0.174352 | 0.3619588 |
|  | 1.5 | 3.5013** | 4.1893** | 0.168648 | 0.3556661 |
|  | 2 | 3.1706** | 3.7235** | 0.15553 | 0.312547 |
| S&P | k = 0.25 | 5.7324** | 5.5284** | 0.12405 | 0.3157433 |
|  | 0.5 | 6.0847** | 5.9749** | 0.138839 | 0.385766 |
|  | 1 | 5.9084** | 5.7912** | 0.144529 | 0.4221734 |
|  | 1.5 | 5.4511** | 5.4709** | 0.142107 | 0.4150753 |
|  | 2 | 4.8572** | 4.7648** | 0.134813 | 0.3655939 |
| **Panel C: Squared Volatility** | | | | | |
| REIT | k = 0.25 | 4.2062** | 4.0009** | 0.162991 | 0.3362795 |
|  | 0.5 | 3.8566** | 4.5319** | 0.174352 | 0.3619588 |
|  | 1 | 3.1706** | 3.7235** | 0.15553 | 0.312547 |
|  | 1.5 | 2.5481** | 2.8687** | 0.129716 | 0.2333424 |
|  | 2 | 2.0718* | 2.2246* | 0.110291 | 0.1758527 |
| S&P | k = 0.25 | 6.0847** | 5.9749** | 0.138839 | 0.385766 |
|  | 0.5 | 5.9084** | 5.7912** | 0.144529 | 0.4221734 |
|  | 1 | 4.8572** | 4.7648** | 0.134813 | 0.3655939 |
|  | 1.5 | 3.6671** | 3.5083** | 0.117991 | 0.2193724 |
|  | 2 | 2.725** | 1.7617 | 0.103408 | 0.1310806 |

Notes: Further details of the long memory tests and parameter estimates are given in the text. The R/S test is the modified R/S statistic (Lo, 1991). The GPH test is the Geweke & Porter-Hudak (1983) non parametric statistic. The R/S $d$ is the R/S long memory parameter. The Perododgram $d$ is the Periodogram long memory parameter. Estimates are given for returns (Panel A) and for Absolute Volatility (Panel B) and Squared Volatility (Panel C) with different power transformations, $k$, * represents significance at 5% level. ** represents significance at 1% level.

**Table 3: Fractionally Integrated GARCH Models for Daily Return Series**

|  | REITs | | S&P 500 | |
|---|---|---|---|---|
|  | Coefficient | p-value | Coefficient | p-value |
| **Panel A: FIGARCH (1,1)** | | | | |
| A | 0.04641*** | 3.66E-06 | 0.02803*** | 2.10E-07 |
| GARCH(1) | 0.50543*** | 1.01E-11 | 0.54397*** | 0.00E+00 |
| ARCH(1) | 0.37822*** | 1.69E-08 | 0.18921*** | 8.22E-15 |
| D | 0.3175*** | 0.00E+00 | 0.39632*** | 0.00E+00 |
| LM (12) | 20.9* | 0.05189 | 7.942 | 0.7896 |
| $Q^2$ (12) | 20.29* | 0.06178 | 20.32* | 0.06127 |
| **Panel B: FIEGARCH (1,1)** | | | | |
| A | -0.24867*** | 0.00E+00 | -0.10651*** | 0.00E+00 |
| GARCH(1) | 0.13449** | 2.72E-02 | 0.452*** | 7.38E-08 |
| ARCH(1) | 0.33099*** | 0.00E+00 | 0.13623*** | 0.00E+00 |
| Leverage | -0.05662*** | 1.41E-08 | -0.0983*** | 0.00E+00 |
| d | 0.59397*** | 0.00E+00 | 0.63067*** | 0.00E+00 |
| LM (12) | 17.94 | 0.1176 | 9.205 | 0.6853 |
| $Q^2$ (12) | 17.61 | 0.1279 | 9.118 | 0.6928 |

Notes: Coefficients and marginal significance levels for the FI(E)GARCH models are presented. A single asterisk denotes statistical significance at the 10%, two denotes statistical significance at the 5% level, while three denotes statistical significance at the 1% level. The FIEGARCH incorporates a leverage variable. The diagnostics are supportive a good fit for both fractionally integrated models. The diagnostic used are the $Q^2$(12) Ljung-Box test on the squared standardised residual series and Engle's (1982) LM test for up to 12$^{th}$ order ARCH effects on the squared standardised returns series.

**Figure 5: Time Series Plots of Daily Volume Series**

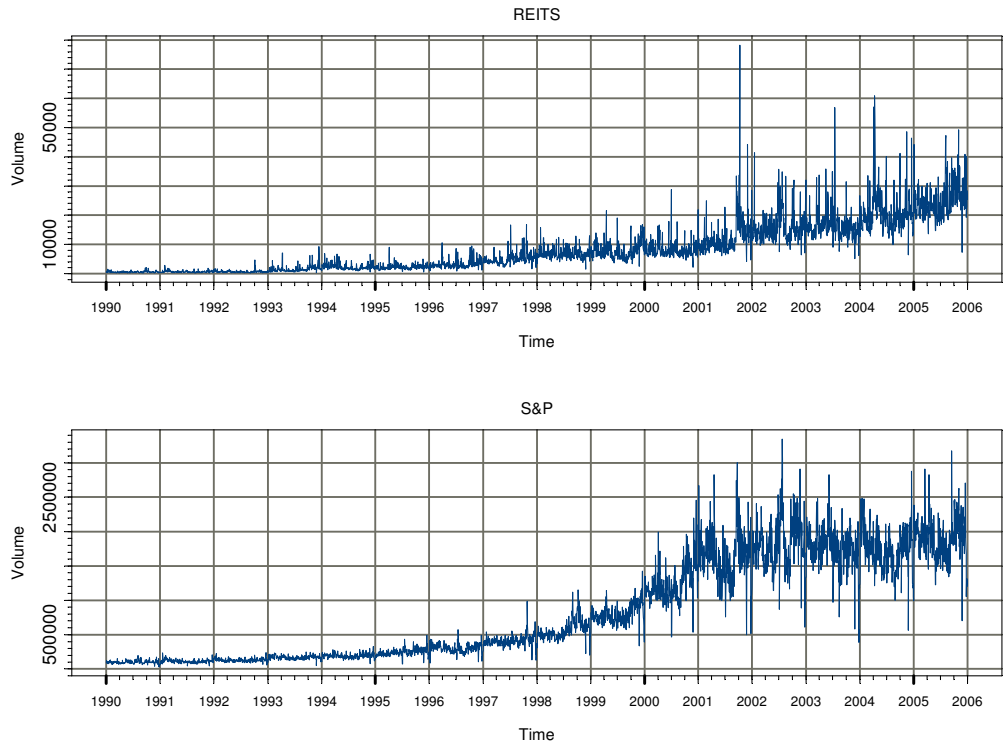

**Figure 6: Time Series Plots of Daily Change in Volume Series**

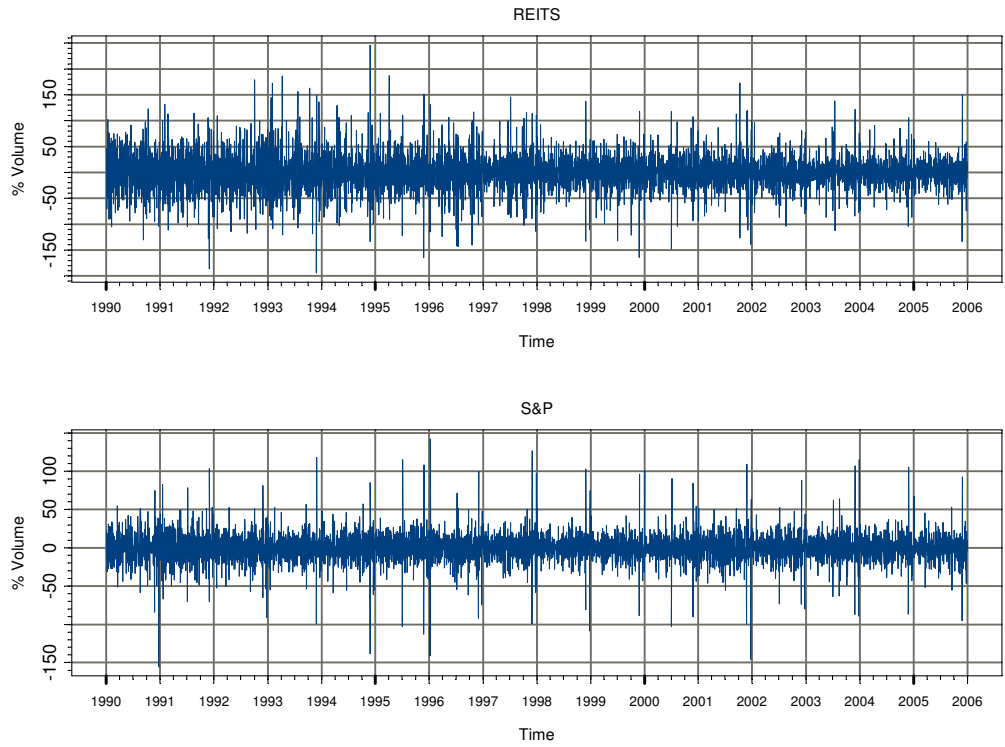

**Table 4: Fractionally Integrated EGARCH Model with Volume**

|  | REITs | | S&P 500 | |
|---|---|---|---|---|
|  | **Coefficient** | **p-value** | **Coefficient** | **p-value** |
| **FIEGARCH (1,1)** | | | | |
| A | -0.14787*** | 0.00E+00 | -0.08081*** | 2.00E-15 |
| GARCH(1) | 0.17908*** | 9.65E-04 | 0.08229* | 8.82E-02 |
| ARCH(1) | 0.19329*** | 0.00E+00 | 0.1001*** | 2.07E-14 |
| Leverage | -0.03489*** | 9.13E-08 | -0.06085*** | 1.45E-13 |
| Volume | 0.01184*** | 0.00E+00 | 0.02012*** | 0.00E+00 |
| d | 0.77339*** | 0.00E+00 | 0.86554*** | 0.00E+00 |
| LM (12) | 20.84* | 0.05277 | 21.21* | 0.04734 |
| $Q^2$ (12) | 20.71* | 0.05477 | 21.95* | 0.03811 |

Notes: Further details of the models are in the text. Model coefficients and associated p values are presented. A single asterisk denotes statistical significance at the 10%, two denotes statistical significance at the 5% level, while three denotes statistical significance at the 1% level. Optimal models are chosen based on Akaike's (AIC) and Schwarz's (BIC) selection criteria. The LM diagnostic determines if there is any ARCH effects remaining in the standardised residuals. The $Q^2$ (12) diagnostic is the Ljung-Box statistic on the squared standardised residuals.

**Figure 7: Time Series Plots of FIEGARCH Daily Conditional Volatility Series**

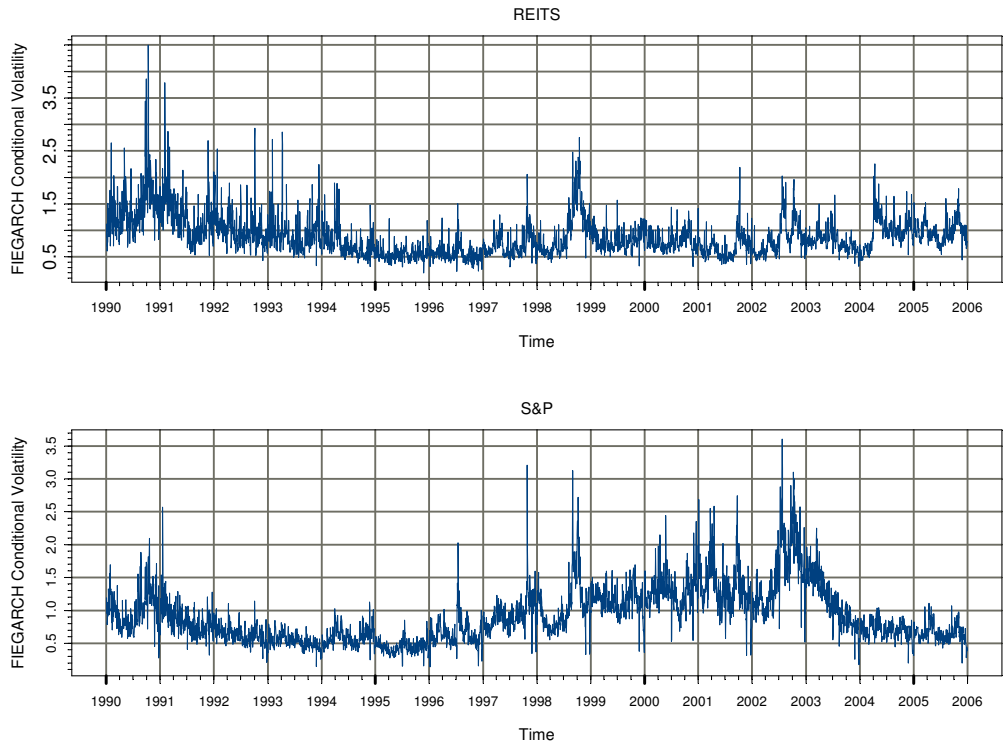

**Endnotes:**

[1] Two recent papers to have examined persistence and mean reversion in REIT and international real estate security returns are Kleiman et al. (2002) and Stevenson (2002a).

[2] As volatility is a latent unobservable variable proxies of volatility such as absolute and squared returns are examined in the literature.

[3] For an excellent treatment of long memory processes see Beran (1994).

[4] Extensions of the Hurst (1951) *R/S* statistic involve replacing the sample standard deviation of the series, *Z*, with the square root of the Newey-West estimate of the long run variance.

[5] One such example of a relatively successful application of standard GARCH models is the application of the APARCH model developed by Ding et al. (1993). The APARCH specification nests seven commonly applied GARCH models. However, the specification has an exponential decline structure that shows strong dependence but is not fully consistent with the long memory decline structure.

6 See Lamoureux & Lastrapes (1990). They find that trading volume reflects the dependence in information flows to the market that feeds in directly into price volatility.

[7] We avoid fitting the FGARCH specification as our exogenous variable, change in trading volume, is not always positive as can be seen from the time series plot and would result in negative conditional variance values.